\documentclass[12pt,a4paper]{article}
\usepackage{a4wide}
\usepackage{latexsym}
\usepackage{epsf}
\usepackage{amssymb}
\makeatletter
\@addtoreset{equation}{section}
\makeatother

\begin{document}

\begin{flushright}
\small
IFT-UAM/CSIC-00-12\\
{\bf hep-th/0003248}\\
March $27$th, $2000$
\normalsize
\end{flushright}

\begin{center}


\vspace{.7cm}

{\LARGE {\bf Supersymmetric Brane-Worlds}}

\vspace{1.2cm}

{\bf\large Natxo Alonso-Alberca},${}^{\spadesuit}$
\footnote{E-mail: {\tt Natxo.Alonso@uam.es}}
{\bf\large Patrick Meessen}${}^{\spadesuit}$
\footnote{E-mail: {\tt Patrick.Meessen@uam.es}}
\vskip 0.3truecm
{\bf\large and Tom\'as Ort\'{\i}n}${}^{\spadesuit\clubsuit}$
\footnote{E-mail: {\tt tomas@leonidas.imaff.csic.es}}
\vskip 1truecm

${}^{\spadesuit}$\ {\it Instituto de F\'{\i}sica Te\'orica, C-XVI,
Universidad Aut\'onoma de Madrid \\
E-28049-Madrid, Spain}

\vskip 0.2cm
${}^{\clubsuit}$\ {\it I.M.A.F.F., C.S.I.C., Calle de Serrano 113 bis\\ 
E-28006-Madrid, Spain}

\vspace{.7cm}


{\bf Abstract}

\end{center}

\begin{quotation}

\small

We present warped metrics which solve Einstein equations
with arbitrary cosmological constants in both in upper and lower
dimensions. When the lower-dimensional metric is the maximally
symmetric one compatible with the chosen value of the cosmological
constant, the upper-dimensional metric is also the maximally symmetric
one and there is maximal unbroken supersymmetry as well.

We then introduce brane sources and find solutions with analogous
properties, except for supersymmetry, which is generically broken in
the orbifolding procedure (one half is preserved in two special
cases), and analyze metric perturbations in these backgrounds

In analogy with the D8-brane we propose an effective
$(\hat{d}-2)$-brane action which acts as a source for the RS solution.
The action consists of a Nambu-Goto piece and a Wess-Zumino
term containing a $(\hat{d}-1)$-form field. It has the standard form
of the action for a BPS extended object, in correspondence with
the supersymmetry preserved by the solution.

\end{quotation}

\newpage

\pagestyle{plain}


\section*{Introduction}

Randall and Sundrum's recent proposal for an alternative to standard
Kaluza-Klein (KK) compactification in Refs.~\cite{kn:RS1,kn:RS2} has
attracted a lot of attention from many quarters: from a
phenomenological point of view, it is a new and fresh proposal to
understand the hierarchy between gauge and gravitational interactions,
while from a purely gravitational point of view it rises many
interesting problems concerning the relation between bulk and brane
gravitational phenomena. In any case, these models provide a new arena
in which one can study new and old problems of Theoretical Physics.

It is worth trying to extend this framework. Here we will present
generalizations of the Randall-Sundrum (RS) scenario which could be
used as alternatives to KK compactification. They are solutions of the
Einstein equations with arbitrary cosmological constant in $\hat{d}$
dimensions and lead to $d=(\hat{d}-1)$-dimensional metrics which solve
the Einstein equations with arbitrary $d$-dimensional cosmological
constant. They have a property which one should require of any
framework with extra dimensions: when the $d$-dimensional metric is
maximally symmetric (and, therefore, is the lower-dimensional vacuum)
the corresponding $\hat{d}$-dimensional metric is also maximally
symmetric (and, therefore, the upper-dimensional vacuum). This holds
in any consistent standard KK compactification: vanishing matter
fields and Minkowski metric in lower dimensions (the $d$-dimensional
vacuum) correspond to Minkowski times a torus metric in upper
dimensions (the $\hat{d}$-dimensional vacuum). The same can be said of
supersymmetry although there are subtleties that in many cases will
make impossible to define lower-dimensional supersymmetry.

In order to exploit these ``bulk'' solutions for dimensional
reduction, we introduce brane sources, find the modified solutions and
study the dynamics of gravitons in the new backgrounds. We also find
the effective gravity actions and Newton constants in lower dimensions
and study supersymmetry on the brane-worlds.

In finding the solutions with branes, we have to allow for
cosmological constants that are piecewise constant functions, a fact
which comes naturally when dualizing it. Therefore we consider gravity
coupled to a volume-form field strength and coupled to a generic
$(\hat{d}-2)$-brane action.


\section{Bulk Solutions}

We are interested in ``warped metrics'' of the form\footnote{We work
  in arbitrary dimension $\hat{d}$ with mostly minus signature. All
  $\hat{d}$-dimensional objects carry hats. We choose
  $x^{\hat{d}-1}\equiv y$ as the spacelike holographic coordinate and
  thus, we split the $\{\hat{x}^{\hat{\mu}}\}=\{x^{\mu},y\}$. Unhatted
  objects are $d\,(=\hat{d}-1)$-dimensional.}

\begin{equation}
\label{eq:ansatz}
d\hat{s}^{2} = a^{2}(y)\, ds^{2} -dy^{2}\, ,
\hspace{1cm}
ds^{2} =  g_{\mu\nu}(x)\, dx^{\mu}dx^{\nu}\, ,
\end{equation}

\noindent solving the equations 

\begin{equation}
\hat{R}^{\hat{\mu}\hat{\nu}} =
\hat{\Lambda} \hat{g}^{\hat{\mu}\hat{\nu}}\, ,
\hspace{1cm}
R^{\mu\nu} = \Lambda g^{\mu\nu}\, ,
\end{equation}

\noindent where $\hat{\Lambda}$ and $\Lambda$ are respectively 
the $\hat{d}$ and $d$-dimensional cosmological constants whose signs
are, in principle, arbitrary. We define $\hat{g}$ and $g$ by

\begin{equation}
\hat{g}^{2} =-\frac{\hat{\Lambda}}{(\hat{d}-1)}\, ,
\hspace{2cm}
g^{2} = -\frac{\Lambda}{(d-1)}\, .
\end{equation}

The solutions fall into two classes:

\begin{equation}
\mathbf{1.\,\,\,\,\hat{g}\neq 0}\hspace{4cm}   
a(y) = {\textstyle\frac{1}{2}}\sqrt{\pm g^{2}/\hat{g}^{2}} 
       \left(e^{\hat{g}y} \pm e^{-\hat{g}y}\right)\, ,
\hspace{3cm}
\end{equation}
  
\noindent where the sign has to be chosen such as to make $a(y)$ real. 
This is always possible except for the case $\hat{g} \in \mathbb{I}\,
,g \in \mathbb{R}$. In the other cases we have\footnote{The case in
  which $a$ is not real can be fixed by Wick-rotating $y$ into a
  timelike coordinate.}, with $g\neq 0$

\begin{equation}
\mathbf{(a)\,\,\,\,\hat{g}\, ,g \in \mathbb{R}}\hspace{5.5cm} 
a = g/\hat{g}\,  \cosh{\hat{g}y}\, ,\hspace{4cm} 
\end{equation}

\begin{equation}
\mathbf{(b)\,\,\,\,\hat{g}\in \mathbb{R}\, , g\in \mathbb{I}}\hspace{4.7cm}
     a = ig/\hat{g}\,  \sinh{\hat{g}y}\, , \hspace{4cm}
\end{equation}

\begin{equation}
\mathbf{(c)\,\,\,\, \hat{g}\, ,g \in \mathbb{I}}\hspace{5.5cm} 
 a = g/\hat{g}\,  \cos{i\hat{g}y}\, .\hspace{4cm}
\end{equation}
  
In this case, the coordinate $y$ naturally lives in a circle of length
$\frac{2\pi}{i\hat{g}}$.

With $g=0$ the only possibility is $\mathbf{\hat{g}\in \mathbb{R}}$
and 

\begin{equation}
a = e^{\pm\hat{g}y}\, .
\end{equation}

\begin{equation}
\mathbf{2.\,\,\,\,\hat{g}=0}\hspace{6cm}    a= igy\, ,\hspace{5cm}
\end{equation}
  
\noindent  which means that we must have $g\in \mathbb{I}$.

The main property of these solutions is that, if $g_{\mu\nu}$ is the
maximally symmetric metric in $d$ dimensions with curvature given by
$\Lambda$, then $\hat{g}_{\hat{\mu}\hat{\nu}}$ is the maximally
symmetric metric with curvature given by $\hat{\Lambda}$. The RS
solution \cite{kn:RS1,kn:RS2} fits in the $\hat{g}\in\mathbb{R}\,
,g=0$ case: if $g_{\mu\nu}=\eta_{\mu\nu}$, we have upstairs (locally)
anti-De Sitter (aDS).  Other possibilities that we are introducing
here are: to have either aDS or DS both upstairs and downstairs, to
have aDS upstairs and DS downstairs and to have Minkowski upstairs and
DS downstairs.  The most interesting options (at least from the
supersymmetry point of view) are the RS solution and the one with
Minkowski upstairs and DS downstairs.

In any dimension, in absence of other fields, the gravitino
supersymmetry transformation law will take the form\footnote{Depending
  on the dimension, we will have one or another kind of minimal
  spinors associated to representations of the gamma matrices with
  special properties. This will never be an issue in what follows and
  our results can be adapted to all the cases of interest.}

\begin{equation}
\delta_{\hat{\epsilon}}\hat{\psi}_{\hat{\mu}}
= \hat{\cal D}_{\hat{\mu}} \hat{\epsilon}\, ,
\end{equation}

\noindent where $\hat{\cal D}_{\hat{\mu}}$ is the aDS 
($\hat{g}\in\mathbb{R}$) or Lorentz ($\hat{g}=0$) covariant
derivative\footnote{Formally we can also consider the DS case
  ($\hat{g}\in\mathbb{I}$). DS supergravities do exist even though
  they are inconsistent as quantum theories.}

\begin{equation}
\hat{\cal D}_{\hat{\mu}} =
\partial_{\hat{\mu}} 
-{\textstyle\frac{1}{4}}\hat{\omega}_{\hat{\mu}}{}^{\hat{a}\hat{b}}
\hat{\gamma}_{\hat{a}\hat{b}}  
-{\textstyle\frac{i}{2}}\hat{g}\hat{\gamma}_{\hat{\mu}}\, .
\end{equation}

Then, the Killing-spinor equation
$\delta_{\hat{\epsilon}}\hat{\psi}_{\hat{\mu}}=0$ has the following
solutions:

  \begin{equation}
\mathbf{1.\,\,\,\,\hat{g}\neq 0}\hspace{2cm}   
\hat{\epsilon} = 
{\textstyle\frac{1}{2}}\left(e^{\hat{g}y/2} +\varphi e^{-\hat{g}y/2}\right) 
\epsilon_{+}
+
{\textstyle\frac{1}{2}}\left(e^{\hat{g}y/2} -\varphi e^{-\hat{g}y/2}\right) 
\epsilon_{-}\, ,\hspace{2cm}
  \end{equation}

\noindent where $\varphi= (g/\hat{g})/|g/\hat{g}|$ and where $\epsilon_{\pm}$
are two spinors that satisfy

\begin{equation}
\left({\cal D}_{\mu} 
\mp {\textstyle\frac{i}{2}}g \gamma_{\mu}\right)\epsilon_{\pm}=0\, ,
\end{equation}

\noindent ${\cal D}_{\mu}$ being the standard Lorentz covariant derivative
and $\gamma_{a}\equiv \hat{\gamma}_{a}$. These equations have maximal number
of solutions when the $d$-dimensional space is maximally symmetric.

\noindent $\mathbf{2.\,\,\,\,\hat{g}=0}$. 
The solution in this case is any $y$-independent spinor
$\hat{\epsilon}$ satisfying

\begin{equation}
\left({\cal D}_{\mu} 
-{\textstyle\frac{i}{2}}g \gamma_{\mu}\right)\hat{\epsilon}=0\, ,
\end{equation}

\noindent where now $\gamma_{a}\equiv\hat{\gamma}_{a}\hat{\gamma}_{y}$.
In this case we had to take $g\in \mathbb{I}$ and thus this is the
$d$-dimensional DS covariant derivative. This equation has a maximal
number of solutions when the $d$-dimensional spacetime is DS.

Observe that, although DS supergravity is inconsistent, any pure
gravity solution of that theory can be considered a warped
compactification of standard (Poincar\'e) supergravity in one
dimension more.

Although we have managed to reduce the $\hat{d}$-dimensional
Killing-spinor equation to a $d$-dimensional-looking Killing-spinor
equation, this does not mean that we have supersymmetry in the
$d$-dimensional space. In the $\hat{g}\neq 0$ case, we cannot have two
different signs for $g$. Keeping only one means keeping either
$\epsilon_{+}$ or $\epsilon_{-}$, but this truncation is only
consistent with $d$-dimensional Lorentz invariance when $g=0$ (the RS
case). On the other hand in the $\hat{g}=0$ it seems that there is no
problem to have DS supersymmetry. The supersymmetry of the RS solution
has also been studied in Refs.~\cite{kn:ABN} and \cite{kn:GP}. We will
make further comments on their results in the next section.


\section{Brane-World Solutions}
\label{sec-BWsol}

Now, mimicking Randall and Sundrum we consider the gravity plus
brane-sources equations

\begin{equation}
\label{eq:eomsources}
\begin{array}{rcl}
\hat{R}^{\hat{\mu}\hat{\nu}} & = & 
\hat{\Lambda}\hat{g}^{\hat{\mu}\hat{\nu}}
-\hat{\chi} \left[g^{\rho\sigma}\delta_{\rho}{}^{\hat{\mu}}
\delta_{\sigma}{}^{\hat{\nu}} 
-{\textstyle\frac{1}{\hat{d}-2}}
\hat{g}^{\hat{\mu}\hat{\nu}} (\hat{g}^{\rho\sigma}\hat{g}_{\rho\sigma})
\right]\sum_{n}T_{n}\delta(y-y_{n})\, ,\\
& & \\
R^{\mu\nu} & = & \Lambda g^{\mu\nu}\, .\\
\end{array}
\end{equation}

Although we write cosmological {\it constants}, we will have to allow
for {\it piecewise constant functions} of $y$. Then, by making
identifications if necessary we can restrict ourselves to a domain in
which they are really constant.

With the same Ansatz for the metric Eq.~(\ref{eq:ansatz}) these
equations reduce to

\begin{equation}
\left\{
\begin{array}{rcl}
0 &=& a^{\prime\prime}+{\displaystyle\frac{\hat{\Lambda}}{\hat{d}-1}}a
+{\displaystyle\frac{2\hat{\chi}}{\hat{d}-2}}a\sum_{n} T_{n}\delta(y-y_{n})\, ,\\
& & \\
0 &=& (a^{\prime})^2+{\displaystyle\frac{\hat{\Lambda}}{\hat{d}-1}}a^2
-{\displaystyle\frac{\Lambda}{d-1}}\, .\\
\end{array}
\right.
\end{equation}

It is straightforward to see that the solutions take now the form

\begin{equation}
\mathbf{1.\,\,\,\,\hat{g},g\neq 0}\hspace{1cm}
   a(y) = {\textstyle\frac{1}{2}} \sqrt{\pm g^{2}/\hat{g}^{2}} 
\left(e^{\sum_{n}c_{n}|y-y_{n}|+C} \pm e^{-\sum_{n}c_{n}|y-y_{n}|-C}\right)\, ,
\hspace{2cm}
  \end{equation}
  
\noindent where $\hat{g}$ and $g$ are defined as before but now $\hat{g}$
takes the value

\begin{equation}
\hat{g} = \sum_{n}c_{n}[2\theta(y-y_{n})-1]\, ,
\end{equation}

\noindent and $g$ is proportional to $\hat{g}$ with an arbitrary 
proportionality constant so $\hat{g}/g$ is a true (purely real or
imaginary) constant. The simultaneously purely real or imaginary
constants $c_{n}$ are given by

\begin{equation}
c_{n}= 
\left.
-\frac{\hat{\chi} T_{n}}{2(\hat{d}-2)}\, 
{\rm tanh}^{\mp 1}\left(\sum_{m}c_{m}|y-y_{m}|+C\right)
\right|_{y=y_{n}}\, .
\end{equation}

\begin{equation}
\mathbf{2.\,\,\,\,\hat{g}\neq 0\, ,g=0}\hspace{3.5cm}
   a(y) = e^{\sum_{n}c_{n}|y-y_{n}|}\, ,\hspace{5cm}
\end{equation}

\noindent where $\hat{g}$ and the simultaneously purely real 
constants $c_{n}$ are given by

\begin{equation}
\hat{g} = \sum_{n}c_{n}[2\theta(y-y_{n})-1]\, ,
\hspace{2cm}
c_{n}= -\frac{\hat{\chi} T_{n}}{2(\hat{d}-2)}\, ,
\end{equation}

\noindent so 

\begin{equation}
\label{eq:RSsol}
   a(y) = e^{-\frac{\hat{\chi}}{2(\hat{d}-2)}\sum_{n} T_{n} |y-y_{n}|}\, .
\end{equation}

\begin{equation}
\mathbf{3.\,\,\,\,\hat{g}=0}\hspace{3cm}
    a= \sum_{n}c_{n}|y-y_{n}|+C\, ,\hspace{5cm}
\end{equation}
  
\noindent  with 

\begin{equation}
g=  \sum_{n}c_{n}[2\theta(y-y_{n})-1]\, ,  
\hspace{2cm}
c_{n} = 
\left.-\frac{\hat{\chi} T_{n}}{2(\hat{d}-2)}
\frac{1}{\sum_{m}c_{m}|y-y_{m}|+C}\right|_{y=y_{n}}\, .
\end{equation}

In general the equations for the constants $c_{n}$ only have solution
if all of them (and, therefore, the tensions $T_{n}$) have the same
sign.  In particular, a system with two branes only has solution if
both branes have the same tension. {\it The exception is the $g=0$
  (RS) case} in which one can get solutions for arbitrary tensions
(Eq.~(\ref{eq:RSsol})).

The problem of finding the different $c_{n}$'s does not show up if one
considers an infinite periodic array of branes and anti-branes with
opposite tensions. We can restrict ourselves to a fundamental region
bounded by two branes or anti-branes with an anti-brane (resp.~brane)
in the middle. The system is mirror symmetric with respect to the
middle (anti-) brane and we can make a further $\mathbb{Z}_{2}$
identification that leaves us with a piece of spacetime bounded by a
brane and an anti-brane in which $\hat{\Lambda}$ and $\Lambda$ are
constant (and in which only one constant $c_{n}$ matters). In these
conditions, taking as fundamental region the interval $y\in
[0,\ell/2]$ with an anti-brane placed at $y=0$ and a brane at
$y=\ell/2$ the warp function $a(y)$ takes the same form as if there
was only one brane in the whole spacetime:

\begin{enumerate}

\item $\mathbf{\hat{g},g\neq 0}$

\begin{equation}
\mathbf{(a)\,\,\,\,\hat{g}\, ,g \in \mathbb{R}} \hspace{1.5cm}
     a = g/\hat{g}\,  \cosh\left( \hat{g}|y|+C \right)\, , \hspace{1.5cm}
\hat{g}=-\frac{ \hat{\chi} T\coth (C) }{\hat{d}-2 }\, .
\end{equation}

\begin{equation}
\mathbf{(b)\,\,\,\,\hat{g}\in \mathbb{R}\, , g\in \mathbb{I}}\hspace{1cm}
a = ig/\hat{g}\,  \sinh\left( \hat{g}|y|+C\right) \, , \hspace{1.3cm}
\hat{g}=-\frac{ \hat{\chi} T \tanh (C)}{\hat{d}-2}\, .
\end{equation}

\begin{equation}
\mathbf{(c)\,\,\,\,\hat{g}\, ,g \in \mathbb{I}}\hspace{1.8cm}
 a = g/\hat{g}\,  \cos\left( i\hat{g}|y|+C \right)\, ,\hspace{1.6cm}
\hat{g}=-i\frac{\hat{\chi} T \coth (C)}{ \hat{d}-2 }\, .
\end{equation}
    
In this case, $\ell$ must be an integer fraction of the period of $y$
i.e.~$\ell=\frac{2\pi}{in\hat{g}}$.

\item $\mathbf{\hat{g}\neq 0\, , g=0}$. $\mathbf{\hat{g}\in
    \mathbb{R}}$
 
    \begin{equation}
     a = e^{\hat{g}|y|}\, ,\hspace{2cm}\hat{g}=-\hat{\chi} T/2\, .
    \end{equation}
    
\item $\mathbf{\hat{g}=0}$

  \begin{equation}
    a= ig|y|+C\, ,\hspace{2cm} g= i\frac{ \hat{\chi} TC }{ \hat{d}-2 }\, .
  \end{equation}

\end{enumerate}

Let us now consider the bulk and world-brane supersymmetry of these
solutions. We can only have supersymmetry on the brane in the RS case
$\hat{g}\neq 0,g=0$ and in the Minkowski-DS case $\hat{g}= 0,g\neq 0$
and imaginary. In these two cases the amount of supersymmetry
preserved depends on the $d$-dimensional (brane) metric $g_{\mu\nu}$.
If it is maximally symmetric, then there will be maximal supersymmetry
on the brane.

Generic branes generically break $\hat{d}$-dimensional bulk
supersymmetry.\footnote{We are not going to include sources in the
supersymmetry transformation rules as in Ref.~\cite{kn:ABN}.  We
think one really needs proper $\kappa$-symmetric brane-sources in
order to study in a fully consistent way the supersymmetric source
problem.} 
However, in these cases, supersymmetry is not broken
{\it locally} in the bulk, in between any pair of branes, since there
the metric has exactly the same form as in the absence of branes.

One may want to have unbroken supersymmetry globally, an not just in
between the branes. First, we need to be able to define the
Killing-spinor equation globally.
In order to do this, we have to allow for a
$\hat{g}$ which is piecewise constant instead of globally constant
(the main characteristic of these branes is that the value of
$\hat{g}$ is different in both sides). We have implicitly accepted
this generalization in this section in order to find the solutions. On
the other hand, one can use a dual formulation in which the
cosmological constant is replaced by a $d$-form potential as in
Ref.~\cite{kn:BRGPT}, an idea which will be investigated in
Sec.~(\ref{sec-braneaction}).
Once we accept this generalization, the necessary condition to have
global unbroken supersymmetry is to be able to match the solutions of
the Killing-spinor equation in both sides of a given brane.  Let us
take, for simplicity, one brane placed at $y=0$. Both $\hat{g}$ and
$g$ change sign across the brane. In the $y>0$ side of the brane, the
solutions of the Killing-spinor equation are those exhibited in the
previous section. In the $y<0$ side of the brane we find solutions of
the same form, where, now, the spinors appearing in the general
solution satisfy
%


%
the same equations but with the sign of $g$ reversed.
We need to set $g=0$ which means that in the second case all
supersymmetry is broken unless we have a trivial solution.

In the first case, it is not enough to have $g=0$ which brings us the
the RS case again. It turns out that we also need to impose the
condition

\begin{equation}
i\hat{\gamma}_{y}\hat{\epsilon}=+\hat{\epsilon}\, ,
\end{equation}

\noindent on the Killing-spinor, which reduces supersymmetry to a half.
This is the same condition we would impose if we were orbifolding the
space between branes.

We would like to stress that our results apply strictly to the cases
we are considering: the infinitely thin branes described by the above
solutions which make the metric across them discontinuous. Thus, our
results do not contradict those of Linde and Kallosh \cite{kn:KL} who
did not study just pure supergravity but included supersymmetric
matter. In that paper the authors tried to find supersymmetric {\it
  thick} domain walls for which the metric is smooth using consistent
superpotentials but did not find any.


\subsection{4-$d$ Action and Newton Constant}

The action from which the equations of motion Eqs.~(\ref{eq:eomsources})
follow has the form

\begin{equation}
\hat{S}= {\textstyle\frac{1}{2{\hat{\chi}}}}
\int{d^{\hat d} \hat{x} \sqrt{|{\hat g}|}\, 
\left[ {\hat R}-({\hat d}-2){\hat \Lambda} \right]}+{\rm branes}\, ,
\end{equation}

\noindent and for $\hat{d}$-dimensional metrics of the warped form
Eq.~(\ref{eq:ansatz}) it reduces to

\begin{equation}
S={\textstyle\frac{1}{2{\hat{\chi}}}}\int{dy\, a^{{\hat d}-3}}
\int{d^{d} x \sqrt{|g|}\, \left[ R-(d-2)\Lambda \right]}\, .
\end{equation}

Comparing, we find that the $d$-dimensional Newton constant $\chi$
is related to the $\hat{d}$-dimensional one $\hat{\chi}$ by 

\begin{equation}
\hat{\chi}/\chi=\int{dy\, a^{{\hat d}-3}}\, .
\end{equation}

Taking $\hat{d}=5$ for definiteness, we can calculate the proportionality
factor in the different cases:

{\bf Case 1.a:} $a=g/{\hat g}\cosh{{\hat g} |y|}$




\begin{equation}
\chi=2\frac{{\hat g}^3/g^2}{\sinh{\left( {\hat g}\ell \right)}
+{\hat g}\ell}{\hat{\chi}}\, .
\end{equation}

{\bf Case 1.b:} $a=ig/{\hat g}\sinh{|{\hat g}| |y|}$


\begin{equation}
\chi=2\frac{|{\hat g}|^3/(ig)^2}{\sinh{\left( {\hat g}\ell \right)}
-{\hat g}\ell}{\hat{\chi}}\, .
\end{equation}

{\bf Case 1.c:} $a=g/{\hat g}\cos{i{\hat g} y}$

In this case the integration limits are $0$ and $2\pi /{i{\hat g}}$:

\begin{equation}
\chi=\frac{\left( i{\hat g} \right)^3}{\pi (ig)^2}{\hat{\chi}}\, .
\end{equation}

{\bf Case 3:} $a=i g |y|+C$

In this case we have:

\begin{equation}
\chi=\frac{3}{2}
\frac{ig}{\left( ig\ell /2+C \right)^3-C^3}{\hat{\chi}}\, .
\end{equation}


\section{Graviton Dynamics}

Expanding the first of Eqs.~(\ref{eq:eomsources}) around a background
which satisfies the same equation one finds the equation of motion for
the perturbation $\hat{h}_{\hat{\mu}\hat{\nu}}$ and using the 
transverse traceless (tt) gauge

\begin{equation}
\hat{\nabla}^{\hat{\mu}}\hat{h}_{\hat{\mu}\hat{\nu}}=\hat{h}=0\, ,
\end{equation}

\noindent where $\hat{h}
=\hat{g}^{\hat{\mu}\hat{\nu}}\hat{h}_{\hat{\mu}\hat{\nu}}$ we
get\footnote{All indices are raised and lowered with the full
  $\hat{d}$-dimensional background metric
  $\hat{g}_{\hat{\mu}\hat{\nu}}$.}

\begin{equation}
 \begin{array}{rcl}
\hat{\nabla}^{2} \hat{h}_{\hat{\mu}\hat{\nu}}
+2\hat{R}_{\hat{\rho}(\hat{\mu}}\hat{h}^{\rho}{}_{\hat{\nu})}
+2\hat{R}^{\hat{\lambda}}{}_{(\hat{\mu}\hat{\nu})}{}^{\hat{\sigma}}
\hat{h}_{\hat{\lambda}\hat{\sigma}}
-2\hat{\Lambda} \hat{h}_{\hat{\mu}\hat{\nu}}-
\hspace{3cm}
& & \\
& & \\
+2\hat{\chi}\left\{\hat{g}^{\rho\sigma}
\left[2\hat{h}_{\rho(\hat{\mu}}\hat{g}_{\hat{\nu})\sigma} 
-\frac{1}{\hat{d}-2}
\left(\hat{h}_{\rho\sigma}\hat{g}_{\hat{\mu}\hat{\nu}}
+\hat{h}_{\hat{\mu}\hat{\nu}}\hat{g}_{\rho\sigma}
\right)\right]
 \right. & & \\
 & & \\
 \left.
-\hat{h}^{\rho\sigma}\left(\hat{g}_{\rho\hat{\mu}}\hat{g}_{\sigma\hat{\nu}}
-\frac{1}{\hat{d}-2}\hat{g}_{\rho\sigma}\hat{g}_{\hat{\mu}\hat{\nu}}
\right)
\right\}\sum_{n}T_{n}\delta(y-y_{n}) & = & 0\, .\\ 
  \end{array}
\end{equation}

\noindent Using now the first of Eqs.~(\ref{eq:eomsources}) to eliminate
$\hat{R}_{\hat{\rho}\hat{\mu}}$ we get 

\begin{equation}
 \begin{array}{rcl}
\hat{\nabla}^{2} \hat{h}_{\hat{\mu}\hat{\nu}}
+2\hat{R}^{\hat{\lambda}}{}_{(\hat{\mu}\hat{\nu})}{}^{\hat{\sigma}}
\hat{h}_{\hat{\lambda}\hat{\sigma}}
+2\hat{\chi}\left[
\left( \hat{g}^{\rho\sigma} \hat{h}_{\rho(\hat{\mu}}
- \hat{h}^{\rho\sigma} \hat{g}_{\rho(\hat{\mu}} \right)
\hat{g}_{\hat{\nu})\sigma}
\right.
& & \\
& & \\
\left.
+\frac{1}{\hat{d}-2}
\left( \hat{h}^{\rho\sigma}\hat{g}_{\rho\sigma}
-\hat{g}^{\rho\sigma}\hat{h}_{\rho\sigma}
\right)\hat{g}_{\hat{\mu}\hat{\nu}}
\right]
\sum_{n}T_{n}\delta(y-y_{n}) & = & 0\, .\\ 
  \end{array}
\end{equation}

\noindent Further, using the RS gauge 

\begin{equation}
\hat{h}_{\mu y}=\hat{h}_{yy}=0\, ,
\end{equation}

\noindent and the fact that the warped general metric Eq.~(\ref{eq:ansatz}) 
is block-diagonal we see that the source terms vanish identically.
The equations for $\hat{h}_{\mu y},\hat{h}_{yy}$ are satisfied
identically and do not become constraints.
The equation for the
remaining piece of the perturbation is

\begin{equation}
  \begin{array}{rcl}
a^{-2}\left[\nabla^{2}\hat{h}_{\mu\nu} 
+2R^{\rho}{}_{(\mu\nu)}{}^{\sigma}\hat{h}_{\rho\sigma}\right] 
-\hat{h}_{\mu\nu}^{\prime\prime} & & \\
& & \\
-(\hat{d}-5) a^{-1}a^{\prime}\hat{h}_{\mu\nu}^{\prime}
+2[(\hat{d}-4)a^{-2}(a^{\prime})^{2} +a^{-1}a^{\prime\prime}]\hat{h}_{\mu\nu} 
& = & 0\, .\\
  \end{array}
\end{equation}

Now we assume that the perturbation can be expanded in RS modes

\begin{equation}
\hat{h}_{\mu\nu}(x,y) =\sum_{\alpha}f_{\alpha}(y)h^{(\alpha)}_{\mu\nu}(x)\, ,
\end{equation}

\noindent of which we only keep the massless one $h^{(0)}_{\mu\nu}
\equiv h_{\mu\nu}$. The sourceless equation of a massless graviton in
a maximally symmetric background, in the tt gauge is

\begin{equation}
\nabla^{2}h_{\mu\nu} 
+2R^{\rho}{}_{(\mu\nu)}{}^{\sigma}h_{\rho\sigma}=0\, ,
\end{equation}

\noindent and, thus, we get for $\hat{h}_{\mu\nu}=f_{0}(y)h_{\mu\nu}$

\begin{equation}
\hat{h}_{\mu\nu}^{\prime\prime}
=(5-\hat{d})a^{-1}a^{\prime}\hat{h}_{\mu\nu}^{\prime}
+2\left[ a^{-1}a^{\prime\prime}
+(\hat{d}-4)a^{-2}(a^{\prime})^2 \right] \hat{h}_{\mu\nu}\, .
\end{equation}

\noindent which in $\hat{d}=5$ is solved by

\begin{equation}
f_{0}(y)=a^{2}(y)\, .
\end{equation}

Depending on the specific solution we can have gravity confinement on
the brane or not. In general, the inclusion of branes and the
orbifolding procedure is necessary to have confinement on just one
brane ($a^{2}$ has more than one maximum in the interval of interest).
The only exception seems to be the RS case. The DS to DS case
($\hat{g},g$ imaginary) deserves special mention because the
holographic coordinate is naturally compact. No branes are needed to
make the graviton wave-function normalizable, although we do need them
if we want to think in terms of confinement.
Some of the general results for a metric of the form (\ref{eq:ansatz})
have also been obtained in Ref.~\cite{kn:CEHS}.


\section{A Brane action for the Randall-Sundrum Scenario}
\label{sec-braneaction}

A constant can be understood as the dual of a volume-form field
strength. A volume-form ({\em i.e.}~a $\hat{d}$-form which we will
denote by $\hat{F}_{(\hat{d})}$) is the field strength of a
$d\equiv(\hat{d}-1)$-form potential $\hat{A}_{(d)}$.  The equation of
motion forces the dual of $\hat{F}_{(\hat{d})}$ to be constant (or,
more generally, piecewise constant). Thus, one can generically
substitute a constant in an action, and, in particular the
cosmological constant, by a $d$-form potential $\hat{A}_{(d)}$. The
canonical example is the rewriting of Romans' massive $10$-dimensional
type~IIA supergravity, which contains a mass parameter $m$, by a
$9$-form Ramond-Ramond potential to which the D8-brane couples
\cite{kn:BRGPT}.

This implies a generalization of the theory since now one can have
solutions in which the value of the cosmological constant is different
in different regions of the spacetime: this is precisely what RS-like
solutions need. The discontinuities are $d$-dimensional topological
defects (domain walls) which act as sources for the $d$-form potential
and can be interpreted as the worldvolumes of $(\hat{d}-2)$-branes
charged under the $d$-form potential (D8-branes in the case of
Ref.~\cite{kn:BRGPT}). A worldvolume action for these branes should,
therefore, contain a Wess-Zumino term: the integral of the pullback of
the $d$-form potential.

All this seems to work very well in the D8-brane case and, in fact,
the rewriting in terms of a 9-form potential proves necessary and even
crucial. It is natural to try something similar here.  Therefore, we
propose an action consisting in a bulk action containing gravity,
$\hat{g}_{\hat{\mu}\hat{\nu}}$, coupled to a $d$-form potential
$\hat{A}_{(d)}$ and a bunch of standard worldvolume actions of
$(\hat{d}-2)$-branes containing the above-mentioned WZ terms with
dynamical coordinate fields $\hat{X}_{n}^{\hat{\mu}}$, {\em i.e.}

\begin{equation}
\begin{array}{rcl}
\hat{S} & = & \frac{1}{2\chi}{\displaystyle\int}d^{\hat{d}}\hat{x}
\sqrt{|\hat{g}|}\, \left[\hat{R} 
+\frac{(-1)^{\hat{d}-2}}{2\cdot \hat{d}!}\hat{F}_{(\hat{d})}^{2}\right]\\
& & \\
& & 
+\sum_{n}
\left\{ 
- \frac{T_{n}}{2}{\displaystyle\int} d^{d}\xi_{n}
\sqrt{|\gamma_{n}|}\, \left[\gamma_{n}^{ij}
\partial_{i}\hat{X}_{n}^{\hat{\mu}}\partial_{j}\hat{X}_{n}^{\hat{\nu}}
\hat{g}_{\hat{\mu}\hat{\nu}}(\hat{X}_{n}) -(\hat{d}-3)\right]
\right.
\\
& & \\
& & 
\left.
+\frac{(-1)^{d} \mu_{n}}{d!}{\displaystyle\int}
d^{d}\xi_{n} 
\hat{A}_{(d)\, \hat{\mu}_{1}\cdots\hat{\mu}_{d}}(\hat{X}_{n})
\partial_{i_{1}}\hat{X}^{\hat{\mu}_{1}} \cdots 
\partial_{i_{d}}\hat{X}^{\hat{\mu}_{d}}
\epsilon^{i_{1}\cdots i_{d}}
\right\}\, .\\
\end{array}
\end{equation}

The field configurations that minimize this action satisfy the
equations of motion for the metric

\begin{equation}
\label{eq:Einstein}
\begin{array}{rcl}
\hat{G}^{\hat{\mu}\hat{\nu}} 
+\frac{(-1)^{(\hat{d}-2)} \mu_{n}}{2\cdot d!}
\left[
\hat{F}_{(\hat{d})}{}^{\hat{\mu}\hat{\rho}_{1}\cdots\hat{\rho}_{d}}
\hat{F}_{(\hat{d})}{}^{\hat{\nu}}{}_{\hat{\rho}_{1}\cdots\hat{\rho}_{d}} -\frac{1}{2\hat{d}}\hat{g}^{\hat{\mu}\hat{\nu}}\hat{F}_{(\hat{d})}^{2}
\right]+\hspace{2cm}& & \\
& & \\
+\frac{\chi}{\sqrt{|\hat{g}|}}
\sum_{n}
T_{n}{\displaystyle\int} d^{d}\xi_{n}
\sqrt{|\gamma_{n}|}\, \gamma_{n}^{ij}
\partial_{i}\hat{X}_{n}^{\hat{\mu}}\partial_{j}\hat{X}_{n}^{\hat{\nu}}
\delta^{\hat{d}}(\hat{x}-\hat{X}_{n}) & = & 0\, ,\\
\end{array}
\end{equation}

\noindent the $d$-form potential

\begin{equation}
\label{eq:potential}
\hat{\nabla}_{\hat{\mu}}
\hat{F}_{(\hat{d})}{}^{\hat{\mu}\hat{\rho}_{1}\cdots\hat{\rho}_{d}}
+{\textstyle\frac{2\chi}{\sqrt{|\hat{g}|}}}
\sum_{n} \mu_{n}{\displaystyle\int} d^{d}\xi_{n}
\epsilon^{i_{1}\cdots i_{d}}
\partial_{i_{1}}\hat{X}^{\hat{\mu}_{1}} \cdots 
\partial_{i_{d}}\hat{X}^{\hat{\mu}_{d}}
\delta^{\hat{d}}(\hat{x}-\hat{X}_{n}) = 0\, ,
\end{equation}

\noindent  the worldvolume metric (after some manipulations)

\begin{equation}
\label{eq:wmetric}
\gamma_{n\, ij} 
-\partial_{i}\hat{X}_{n}^{\hat{\mu}}\partial_{j}\hat{X}_{n}^{\hat{\nu}}
\hat{g}_{\hat{\mu}\hat{\nu}}(\hat{X}_{n})\, =\, 0\, ,
\end{equation}

\noindent and the (coordinate) worldvolume scalars

\begin{equation}
\label{eq:wscalars}
  \begin{array}{rcl}
\nabla^{2}(\gamma)\hat{X}_{n}^{\hat{\mu}} 
+\hat{\Gamma}_{\hat{\rho}\hat{\sigma}}{}^{\hat{\mu}}(\hat{g})
\partial_{i}\hat{X}_{n}^{\hat{\rho}}\partial_{j}\hat{X}_{n}^{\hat{\sigma}}
\gamma_{n}^{ij} +\hspace{7cm}& & \\
& & \\
+\frac{(-1)^{d} \mu_{n}}{T_{n}d!\sqrt{|\gamma_{n}|}}
\hat{F}_{(\hat{d})}{}^{\hat{\mu}}{}_{\hat{\rho}_{1}\cdots\hat{\rho}_{d}}
\partial_{i_{1}}\hat{X}^{\hat{\rho}_{1}} \cdots 
\partial_{i_{d}}\hat{X}^{\hat{\rho}_{d}}
\epsilon^{i_{1}\cdots i_{d}} & = & 0\, .\\
\end{array}
\end{equation}

Eq.~(\ref{eq:wmetric}) simply states that the worldvolume metrics are
those induced on the worldvolumes by the embedding coordinates
$\hat{X}_{n}^{\hat{\mu}}$.  Using worldvolume reparametrization
invariance we can set $d$ coordinates (static gauge) to the values
$\hat{X}^{\mu}_{n} = \delta^{\mu}_{i}\xi_{n}^{i}$.  Furthermore, our
Ansatz for the remaining coordinate is

\begin{equation}
\label{eq:ansatzwscalar}
\hat{X}^{d}_{n}\equiv Y_{n} = y_{n}\, ,
\end{equation}

\noindent where the $y_{n}$'s are constants. We can
perform the volume integrals in
Eqs.~(\ref{eq:Einstein},\ref{eq:potential}) leaving only 1-dimensional
delta functions $\delta(y-y_{n})$.  Also, this implies for the
worldvolume metrics (identifying worldvolume and $d$-dimensional
spacetime indices)

\begin{equation}
\gamma_{n\, \mu\nu} =  \hat{g}_{\mu\nu} = a^{2}(y_{n})g_{\mu\nu}\, .
\end{equation}

For the potential we have 

\begin{equation}
\hat{A}_{(d)\, \mu_{1}\cdots\mu_{d}} =
c a^{d} \frac{\epsilon_{\mu_{1}\cdots\mu_{d}}}{\sqrt{|g|}}\, , 
\Rightarrow
\hat{F}_{(\hat{d})\, \underline{y} \mu_{1}\cdots\mu_{d}} =
c d a^{d} \log^{\prime}{a}\, 
\frac{\epsilon_{\mu_{1}\cdots\mu_{d}}}{\sqrt{|g|}}\, ,
\end{equation}

\noindent where $c$ is a constant to be found and $\epsilon$ is 
the $d$-dimensional Levi-Civit\`a tensor calculated with the
$d$-dimensional metric. 

Let us solve Eqs.~(\ref{eq:wscalars}).  The equations for the
$\hat{X}^{\mu}_{n}$'s are automatically solved. $\hat{F}_{(\hat{d})}$
only contributes to the equations of the $Y_{n}$'s, which are solved
for

\begin{equation}
c=-T_{n}/\mu_{n}\equiv T/\mu\, .
\end{equation}

This implies that all the quotients $T_{n}/\mu_{n}$ must have the same
value, which is a characteristic of BPS objects. Observe that the
$\mu_{n}$'s cannot vanish: had we tried uncharged brane sources we
would have never succeeded.

The equation for the potential becomes

\begin{equation}
T/\mu d\log^{\prime\prime}{a} 
+2\chi\sum_{n}\mu_{n}\delta(y-y_{n})=0\, ,
\end{equation}

\noindent which is solved by a warp factor of the RS type

\begin{equation}
a=e^{-\frac{\chi \mu/T}{d}\sum_{n}\mu_{n}|y-y_{n}|}\, .  
\end{equation}
 
To solve the Einstein equations we first calculate the energy-momentum
tensor of the form potential, {\em i.e.}

\begin{equation}
\hat{F}_{(\hat{d})}{}^{\hat{\mu}\hat{\rho}_{1}\cdots\hat{\rho}_{d}}
\hat{F}_{(\hat{d})}{}^{\hat{\nu}}{}_{\hat{\rho}_{1}\cdots\hat{\rho}_{d}} 
-{\textstyle\frac{1}{2\hat{d}}}
\hat{g}^{\hat{\mu}\hat{\nu}}\hat{F}_{(\hat{d})}^{2}
=\, \left({\textstyle\frac{\chi}{2}}
\sum_{n}\mu_{n}[\theta(y-y_{n})-1] \right)^{2}
\hat{g}^{\hat{\mu}\hat{\nu}}\, ,
\end{equation}

\noindent which describes a piecewise cosmological constant. 
For only two branes with opposite tensions the Einstein equation is
exactly the one in Ref.~\cite{kn:RS1} in the intervals in which the
cosmological constant is constant and therefore admits the same
solutions. In fact, assuming that the $d$-dimensional metric is
Ricci-flat $R_{\mu\nu}=0$, the Einstein equations are solved by the
above warp factor if $(\mu/T)^{2} = {\textstyle\frac{1}{2}}
d/(\hat{d}-2)$ which implies, as in Sec.~(\ref{sec-BWsol}) and
Refs.~\cite{kn:RS1}, that
 
\begin{equation}
a=e^{-\frac{\chi}{(\hat{d}-2)}\sum_{n}T_{n}|y-y_{n}|}\, .
\end{equation}


\section{Conclusions}
\label{sec-conclusions}

We have explored general solutions with warped metrics with and
without branes and we have studied their supersymmetry properties and
the effective theories on the branes, including supersymmetry.

Two cases are singled out by supersymmetry considerations: the
well-known RS case and the case in which the total spacetime is
Minkowski and on the brane one has DS spacetime. The brane breaks a
half of the available supersymmetry in the RS case, a result also
obtained in Ref.~\cite{kn:ABN}, while in the last case a brane seems
to break completely the bulk supersymmetry although one can still
speak of (DS) supersymmetry on the brane-world with its known
problems.

In analogy with the D8-brane effective action, we considered a
formulation of the problem in terms of a dynamical $\hat{d}-2$ brane
with a $\hat{d}-1$ form potential.

\section*{Acknowledgments}
We would like to thank Bert Janssen and Pedro Silva for many
useful conversations. This work was partially
supported by the E.U. TMR program FMRX-CT96-0012
 and by the Spanish grant AEN96-1655.


\begin{thebibliography}{99}




\bibitem{kn:RS2} L.~Randall and R.~Sundrum,
                 {\it Phys.~Rev.~Lett.}~{\bf 83} (1999) 4690-4693.

\bibitem{kn:RS1}  L.~Randall and R.~Sundrum,
                  {\it Phys.~Rev.~Lett.}~{\bf 83} (1999) 3370-3373.

\bibitem{kn:ABN} R.~Altendorfer, J.~Bagger and D.~Nemeschansky,
                 {\tt hep-th/0003117}.

\bibitem{kn:GP} T.~Gherghetta and A.~Pomarol,
                {\tt hep-ph/0003129}.

\bibitem{kn:KL} R.~Kallosh and A.~Linde,
                {\it JHEP} {\bf 0002} (2000) 005.
                 

\bibitem{kn:BRGPT} E.~Bergshoeff, M.~de Roo, M.B.~Green,
                   G.~Papadopoulos and P.K.~Townsend,
                   {\it Nucl.~Phys.}~{\bf B470} (1996) 113-135.

\bibitem{kn:CEHS} C.~Csaki, J.~Erlich, T.J.~Hollowood and Y.~Shirman,
                  {\tt hep-th/0001033}.


\end{thebibliography}
\end{document}